\documentclass[12pt]{article}
\usepackage{aaspp4}
\begin{document}
\title{Lopsidedness in Early Type Disk Galaxies}
\author{Gregory Rudnick, Hans-Walter Rix\altaffilmark{1} \altaffiltext{1}{Alfred P. Sloan Fellow} }
\affil{Steward Observatory, The University of Arizona}

\begin{abstract}
We quantify the mean asymmetry of 54 face-on, early type disk galaxies
 (S0 to Sab) using the amplitude of the $m=1$ azimuthal Fourier
 component of the R-band surface brightness.  We find that the median
 lopsidedness, $\langle A_1/A_0\rangle$, of our sample is 0.11 and
 that the most lopsided $20\%$ of our galaxies have $\langle
 A_1/A_0\rangle\geq0.19$.  Asymmetries in early type disks appear to
 be of similar frequency and strength as in late type disk
 galaxies (Zaritsky and Rix 1997.)  We have observed our early type
 disks in a bandpass (R-Band) in which the light is dominated by stars
 with ages greater than $10^9$ yrs, and therefore are seeing azimuthal
 asymmetries in the stellar {\it mass} distribution.  The similar
 degree of lopsidedness seen in disks of very different star formation
 rates indicates that the lopsidedness in all galactic disks is
 primarily due to azimuthal mass asymmetries.  Hence, $20\%$ of all
 disk galaxies (regardless of Hubble Type) have azimuthal asymmetries,
 $\langle A_1/A_0\rangle\geq0.19$, in their stellar disk mass
 distribution, confirming lopsidedness as a dynamical phenomenon.
\end{abstract}


\section{Introduction}
Just as a human face may mirror traumatic events in its past, so may a
 galaxy's structure reflect its dynamical history.  In this paper we
 explore one potential probe of the dynamical past of disk galaxies,
 their {\it lopsidedness}.  In the context of this paper, lopsidedness
 is defined as a significant bulk asymmetry in the stellar mass
 distribution of a galactic disk.  For nearly face-on galaxies, these
 asymmetries manifest themselves observationally as a shifting of the
 outer isophotes of the stellar light.

Baldwin et al. (1980) were the first to point out that the HI profiles
 of spiral galaxies are frequently lopsided, and suggested that this
 lopsidedness may stem from weak interactions in the galaxy's past, or
 from lopsided orbits.  Richter and Sancisi (1994) using a much larger
 galaxy sample to study the frequency of lopsidedness, concluded that
 over 50$\%$ of all disk galaxies had significant asymmetries in their
 HI profiles.  These HI asymmetries are present both in the projected
 flux distributions and in velocity space.

A systematic attempt to examine the frequency of asymmetries in the
 stellar light of galaxies has been made by Rix and Zaritsky, 1995 and
 Zaritsky and Rix, 1997 (hereafter RZ95 and ZR97) and a recent optical
 study of symmetry in nearby disk galaxies has been carried out by
 Conselice (1998.)  Using near-IR photometry of face-on galaxies, ZR97
 found that about one fifth of all late type spiral galaxies exhibited
 significant lopsidedness.  They defined lopsidedness via the
 amplitude of the $m=1$ Fourier component of the azimuthal
 decomposition of the galaxy's surface brightness.  It is possible to
 construct self-consistent asymmetric disk models (Syer and Tremaine,
 1996) and some N-body work has shown that lopsided instabilities may
 occur in the presence of retrograde orbits (Zang and Hohl, 1978;
 Sellwood and Merritt, 1994) or in the absence of a massive halo
 (Sellwood, 1985).  These authors also find that co-rotating galaxies
 with massive halos are stable against intrinsic $m=1$ instabilities.

Since intrinsic disk instabilities in spirals appear not to be the
 dominant cause of lopsidedness, external potential perturbations are
 the most plausible candidate mechanism.  N-body simulations (Walker
 et al. 1996; ZR97) have shown that the merger of a small galaxy with
 a large disk galaxy can produce the type and degree of asymmetries
 found in RZ95 and ZR97, if their mass ratio is $\sim 1/10$.  Recent
 N-body work by Levine and Sparke (1998) shows that lopsided modes may
 persist in disk galaxies if the disks are off-center with respect to
 a massive host halo with a large, flat core.  They propose that a
 galaxy may become lopsided if it accretes enough matter to push it
 away from the center of the halo.  It stands to reason that these
 asymmetries may also be produced by weak, non-cataclysmic encounters
 with other galaxies (Barnes, Hernquist 1992 and references therein.)
 Lifetimes of these asymmetries can be estimated from phase mixing
 (Baldwin et al. 1980; RZ95) or through analysis of N-Body simulations
 (Walker et al. 1996; ZR97); both estimates point to lifetimes of 5-10
 rotation periods, or $1\sim$Gyr.

For interpreting lopsidedness it is crucial to know whether
 the observable asymmetries in the stellar {\it light} reflect
 asymmetries in the stellar {\it mass}.  Late type disk galaxies (such
 as those used in RZ95 and ZR97) with relatively high star formation
 rates contain many young, bright stars.  As we will show later, in
 Sbc galaxies, up to 25$\%$ of the light in the light in the R-Band
 can come from stars younger than 0.1 Gyrs (OB stars.)  An asymmetric
 distribution of such a luminous, young population could result in a
 high observed asymmetry, even if the underlying mass structure were
 in fact close to symmetric.  To avoid this potential pitfall, we
 assembled a sample of disk galaxies with low star formation rates,
 those limited to Hubble Types from from S0 to Sab (Kennicutt et
 al. 1994), with the aim of comparing the incidence of lopsidedness in
 this new sample to that in RZ95,ZR97.  If significant asymmetries are
 similarly frequent, we may conclude that they are in fact dynamical
 in origin and are not merely azimuthal variations in the
 mass-to-light ratios within the galaxies.

The layout of the paper is as follows.  In \S2 we will discuss the
 sample selection, observations and reduction techniques; In \S3 we
 will describe the data analysis methods.  The results, including the
 frequency of lopsidedness, are presented and discussed in \S4.  In
 \S5 we will present our conclusions as well as the directions for
 future follow-up work.

\section{The Data}
\subsection{Sample Selection $\&$ Observations}
Our sample is comprised of 60 galaxies taken from the RC3 catalogue
 (De Vaucouleurs, 1991), selected according to the following criteria:
 m$_B\leq$14, cz$\leq$10,000 km/s, $b/a \leq$0.65 ($49^\circ \leq i
 \leq0^\circ$), 0$\leq$T$\leq$2, and a maximum diameter of 4$'$.  The
 median diameter of the galaxies on our sample was $2.3'$.  The
 magnitude and redshift limits were chosen to minimize the required
 exposure times, while the axis ratio of the galaxies was constrained
 to reduce projection effects in our analysis.  For reasons explained
 in \S{3.2}, only 54 of these galaxies were included in our final
 sample (see Table 1.)

The galaxies were observed on three runs at the 2.3-meter Bok reflector
 on Kitt Peak (April 15-17th, 1996) and at the 1.5-meter reflector on
 Mt. Bigelow (November 10-13th, 1996 and April 9-14th, 1997).  For
 these three runs, the CCD pixel scales were 0.3$''/{\rm pix}$,
 0.9$''/{\rm pix}$ and 0.4$''/{\rm pix}$, respectively, with
 fields-of-view of 3'$\times$2', 9.8'$\times$9.3' and
 3.3'$\times$3.3'.  The median seeing at the 2.3-meter was 1.5'' while
 the median seeing during the November and April runs on the 60-inch
 were 1.7'' and 1.4'' respectively.

A Nearly-Mould R-band filter ($\lambda_{center}=650nm$) was used at
 the 2.3-meter.  The filters used for the November and April 60-inch
 runs were respectively a Mould R-band filter with
 $\lambda_{center}=650nm$ and a Kron-Cousins R-band filter with
 $\lambda_{center}=650nm$.

In order to emphasize the light coming from stellar populations that
 trace the stellar mass distribution (old stars), observations were
 previously carried out in the near-IR (K and I bands, RZ95; ZR97.)
 In this paper we study galaxies with lower star formation rates;
 where we can expect the R-band light to be a sufficient tracer of
 phase-mixed stellar populations.  Using simple models (see RZ95,
 Appendix B) for the star formation history of Sa galaxies (Kennicutt
 et al. 1994) we determined that the light from stars younger than
 $10^8$ years ($t\leq t_{orbit}$) typically contributes only $4\%$ to
 the total light in the R-Band, while stars younger than $10^9$
 ($t\leq10t_{orbit}$) years contribute $12\%$ of the total R-Band
 light.  Going to longer wavelengths (e.g. I) would slightly decrease
 the contribution from younger stars but this gain is not worth the
 increased observing expense.  In contrast, a galaxy with the star
 formation history of a stereotypical Sbc (Kennicutt et al. 1994),
 would lead to $24\%$ of the R-Band light originating from stars
 younger than $10^8$ years and $50\%$ from stars younger than $10^9$
 years.

\subsection{Reduction}

The initial image reduction was carried out with standard
 IRAF\renewcommand{\thefootnote}{\fnsymbol{footnote}} \footnote{IRAF
 is distributed by the National Optical Astronomical Observatories,
 which are operated by AURA, Inc. under contract to the NSF.}
 routines.  The 2.3-meter images were flat-fielded with a combination of
 dome flats and twilight flats; the 1.5-meter images were flattened with
 twilight and smoothed night sky flats.  For the 2.3-meter images, the
 dome flats were used to remove the small scale variations, while the
 sky flats (smoothed) were used to remove the large scale variations.
 At the 1.5-meter the smoothed night sky and twilight images were all
 combined into one flat which accounted for both the small and large
 scale variations.  The quality of the large-scale flat-fielding was
 estimated from the variance of the median sky level at the four
 corners of the images.  The large diameter of the galaxies compared
 to the field of view at the 2.3-meter, precluded such an estimate in
 many cases.  Whenever the flat-field quality could be checked
 however, it was found to be $1-2\%$.  At the 1.5-meter, the larger
 field of view allowed a more accurate determination of the flat-field
 quality and the images typically were flattened to within $0.5\%$.
 The superior flat-field quality of the 1.5-meter data may be due to the
 high S$/$N night sky flats.

Point sources in the images were selected using {\it daofind} and
 surrounding pixels were excised in the subsequent analysis to a
 radius where the stellar point spread function brightness had
 declined to the level of the sky.

\section{Data Analysis}
\subsection{Fourier Decomposition}
There is no unique measure of asymmetry in galaxies.  Here, we use an
 approach that provides information not only about the magnitude of
 the disk asymmetry, but also about its angular phase and its radial
 dependence.  Although we only analyze the radial dependence of the
 asymmetry, we use the angular dependence in our visual verification
 of lopsided structure.  Fourier decomposition in polar coordinates
 has been employed previously (RZ95 and ZR97 and references therein),
 and we will not describe it in detail.  Briefly, we center a polar
 coordinate grid on the galaxy nucleus, with 16 azimuthal and 45
 radial bins that extend logarithmically from 2.5 to 180 pixels.  For
 our April, 1996 run at the 2.3-meter and our November, 1996 and April,
 1997 runs at the 1.5-meter, this radial outer limit corresponds
 respectively to 0.9$'$, 2.7$'$ and 1.2$'$.  By these limits, our
 $S/N$ had already declined below the level required to perform a
 successful determination of lopsidedness.  The nucleus is identified
 with the brightest point in a galaxy's center; slight deviations from
 this do not affect significantly the subsequent analysis of
 asymmetries in the outer parts (RZ95.)

Specifically, we decompose the image intensity in an annulus of mean
 radius $R_{m}$ into the form: \begin{equation} I(R_m,\phi) = a_o
 \times \Biggl( 1 + \sum_{j=1}^N a_j e^{-i[j(\phi_j-\phi_j^o)]
 }\Biggr)\end{equation} Where for each radius, $\left|a_o\right|(R)$
 is the average luminosity, $\left|a_1\right|(R)$ describes the
 lopsidedness and $\left|a_2\right|(R)$ is the bisymmetric fluctuation
 amplitude (arising from ellipticity, projection effects, etc.)
 Likewise, $\phi_j^o$ is the position angle of each component $a_j$.
 Although asymmetries may put power into any odd $a_j$, most of the
 power is usually concentrated in the $a_1$ term (RZ95.)  We define
 the luminosity normalized quantities $A_j \equiv a_j/a_0$.

In addition to the Fourier decomposition, we fit a bulge-disk profile
 to the galaxy images (using a $R^{1/4}$ bulge and exponential disk
 law.)  We calculate the mean asymmetry, $\langle A _1\rangle$, for
 each galaxy by taking the radial average of the asymmetry from 1.5 to
 2.5 disk scale lengths weighted by the errors as described in \S3.1.
 The radial limits encompass $26\%$ of the disk light and straddle the
 half light radius of the disk.  Averaging over a range in radii
 reduces the effect of isolated asymmetric peaks on our analysis and
 also gives us a global measure of the lopsidedness for each galaxy in
 our sample.

Because the $A_j$'s are positive definite in the presence of errors
 $\Delta\tilde{A}_1$, the expectation value of the $A_1$ measurement
 will be higher than the true value $\tilde{A}_1$.  The measured value
 at each radius is given by $A_1^2= \tilde{A} _1^2+\Delta A_1^2$.  In
 ZR97, this correction was not included in their determination of
 $\langle A_1 \rangle $.  We have examined the effect that this
 omission will have on our comparison with ZR97, and it was found that
 the difference between $\langle \tilde{A}_1\rangle $ and
 $\langle A_1\rangle $ in our sample is about $1\%$.  Therefore, our
 choice to use $\langle \tilde{A}_1\rangle $ instead of $\langle
 A_1\rangle $ will not effect our comparison.

\subsection{Radial Dependence of the Asymmetries}

In Fig. 1, we plot the median $\tilde{A}_1(L(\leq R)/L(tot))$ profiles
 of the most lopsided $20\%$ of our galaxies and the least lopsided
 $80\%$ of our galaxies with $R_{max}\geq 2.5R_{exp}$ ($L(\leq
 R)/L(tot)\geq0.71$.)  The centers of almost all the galaxies are very
 symmetric regardless of the asymmetry in their outer parts.  Therefor
 using the inner portions of galaxies ($L(\leq R)/L(tot) \leq 0.2$)
 will not aid us in determining the $\langle \tilde{A}_1\rangle $ of
 the bulk disk.  The outer radial limit of $2.5R/R_{exp}$ was chosen
 due to $S/N$ constraints.

The decrease in the value of $\tilde{A}_1(L(\leq R)/L(tot))_{lopsided}$
 for $L(\leq R)/L(tot)\geq0.71$ ($R/R_{exp} \geq 2.5$) is due to the
 fact that two of our more lopsided galaxies only have measurements
 out to $R\approx 2.5R_{exp}$.  Because we are missing two highly
 lopsided galaxies for $R\geq 2.5 R_{exp}$, the median lopsidedness of
 the sample drops at these radii.  Note, that if most of the
 asymmetric galaxies were classified as being lopsided because of
 isolated peaks at some random radius, $1.5R_{exp}\leq R
 \leq2.5R_{exp}$, in their $\tilde{A}_1(R)$ profiles, the two median
 plots would look very similar.

Examining the position angle of the asymmetries, we find that there is
 very little variation with radius.  This implies that lopsidedness
 cannot be attributed to the existence of a wound one-armed spiral.

\subsection{Errors}
The analysis program (RZ95,ZR97) uses Poisson statistics and the
 quality of the flat-field to determine the error in $A_1(R)$.  Beyond
 $R\approx R_{exp}$, the errors in measuring ${A}_1(R)$ are dominated
 by the quality of the flat-fielding, $\delta I${\it counts}.  We
 extended our measurements radially until the flux of the galaxy
 dropped to three times the flat-field error, $I_{galaxy} =
 3\times\delta I$.  For an average sky surface brightness,
 $\mu_{sky}=21.5 mag/asec^2$ and an average flat-field quality,
 $\delta I/I_{sky}=1\%$, our typical limiting surface brightness was
 $\mu_{galaxy}=25.3mag/asec^2$.  An underestimate of the flat-field
 errors will lead to spurious asymmetries in the galaxy's outer
 regions.  To explore this potential problem, we examined all galaxies
 which showed significant lopsidedness in their outer regions and
 verified both that the indicated lopsidedness had a visible
 counterpart, and that the flat field quality was correctly
 determined.  None of our measurements of significant lopsidedness
 were found to be caused by these flat-field errors.

By re-analyzing the images with a coarse binning of of 25 radial and
 10 azimuthal bins, we found that the $R_{exp}$ and $\langle
 \tilde{A}_1\rangle $ estimates depend only negligibly on the grid
 parameters.

The analysis of images taken on consecutive nights with identical
 instruments was found with few exceptions to show excellent
 agreement.  In a few cases, the images taken at different telescopes
 with differing instruments exhibited disagreement.  In all but one of
 these cases (NGC~4580), the discrepancies in
 $\langle\tilde{A}_1\rangle$ were accounted for by the known errors
 and usually resulted from differing values for the fitted disk scale
 lengths, or from slight variations in the $A_1(R)$ profiles.  Only
 for NGC~4580, were we unable to account for the differences and so
 discarded it from the sample.

Further, 6 of the 60 objects imaged (not including NGC~4580) were
 eliminated because their radial measurements did not extend to $R\geq
 2.5R_{exp}$.

In the end, the error $\Delta\langle \tilde{A}_1\rangle $ for each
 image is comprised of three different elements added in quadrature,
 i) the statistical error of the mean ii) the variance of the
 distribution as measured between 1.5 and $2.5 R_{exp}$ iii) the
 systematic error (0.035) derived by taking the average difference in
 $\langle \tilde{A}_1\rangle $ between multiple images of the same
 objects.

\section{Results and Discussion}
\subsection{Wavelength Dependence of Lopsidedness}
To illustrate the wavelength dependence of lopsidedness, we obtained
 UBVRI images of 3 late type disk galaxies, known to be quite
 asymmetric in the near-IR from ZR97; the characteristics for these
 galaxies are given in Table 2.

As is shown in Fig. 2, the wavelength of the observation does affect
 the measured lopsidedness, with I-Band measurements giving the lowest
 values of $\tilde{A}_1(R)$.  This is consistent with the prediction
 of stellar population models that much of the U (and B) light may
 come from very young, and hence not yet phase mixed, stars.  Because
 the V,R and I-bands all have very similar $\tilde{A}_1(R)$ profiles,
 we can assume that the V and R-bands both sample the mass
 distribution with comparable accuracy as I.  This implies that our
 method loses no reliability by going to shorter wavelengths than
 those used by ZR97.

\subsubsection{Is Lopsidedness Caused by Dust}
In highly inclined disk galaxies, dust can dim the near side of the
galaxy more than the far side, leading to a non-zero m=1 amplitude.
Using plausible estimates for the optical depth of an SO/Sa/Sab galaxy
($0.5\leq \tau_V{\rm (face-on)} \leq 1.0$), the magnitude of such an
asymmetry in our face-on sample ($\langle i \rangle = 37^\circ$) is at
most $\sim 3 \%$ (Byun et al., 1994.)  Such global dust asymmetries
would manifest themselves as a correlation between the inclination
angle and the frequency of lopsidedness.  A K-S test shows that the
distribution of lopsidedness among the most inclined third of our
sample galaxies has a $96\%$ probability of being drawn from the same
distribution as the least inclined two-thirds.

As dust produces an asymmetry along the minor axis, it should induce a
correlation between $\phi_1^o$ and $\phi_2^o$.  A K-S test reveals
that the cumulative distribution of $\phi_{diff} =|\langle
\phi_2^o \rangle - \langle \phi_1^o \rangle|$ has a $96\%$ chance of
being drawn from a random distribution, as expected for intrinsic
lopsidedness.

 If dust were the primary cause of lopsidedness, the trend of
decreasing asymmetry with increasing wavelength, should continue to
our reddest band.  As Figure 2 shows however, the asymmetry profiles
for V,R and I are almost identical.

\subsection{Frequency of Lopsidedness}
To determine the percentage of galaxies which are ``significantly''
 lopsided, a lopsidedness threshold must be chosen (RZ95,ZR97).
 Alternatively, we can simply use the cumulative distribution of
 $\langle \tilde{A}_1\rangle$.  This distribution is plotted in Figure
 3, and shows that the majority of galaxies are quite symmetric.  The
 median value for $\langle \tilde{A}_1\rangle$ in our sample is 0.11.
 $20\%$ of our galaxies have $\langle \tilde{A}_1\rangle \geq 0.19$.

To compare the frequency of lopsidedness in ZR97 to that in our
 sample, we must understand any systematic differences in the
 reduction or analysis of our different data.  ZR97 used a simpler
 approach to eliminate superimposed foreground stars causing a small
 artificial increase of 0.03 in the measured lopsidedness.  To
 compensate for this, we subtracted 0.03 from the $\langle
 \tilde{A}_1\rangle$ measures of ZR97.  With this correction, the
 median $\langle \tilde{A}_1\rangle$ in ZR97 is 0.13.  Likewise,
 $20\%$ of their galaxies have $\langle \tilde{A}_1\rangle \geq
 0.19$.

Typical errors in the measurement of $\langle \tilde{A}_1\rangle$
 (0.02 for ZR97 and 0.07 for our sample) are similar to the difference
 between the median values and between the $20\%$ points and so the
 frequency of lopsidedness in our sample is indistinguishable to that
 of ZR97.

The most direct determination of the frequency of lopsidedness would
 be achieved with a volume limited sample.  In a magnitude limited
 sample such as ours however, there may be ambiguities because weak,
 tidal interactions may cause both lopsidedness, and increased star
 formation rates (Mihos, Hernquist 1994; Hernquist, Mihos 1995) The
 volume sampled by lopsided galaxies would hence be larger than that
 sampled by symmetric galaxies, leading us to overestimate the actual
 frequency.  However, the link between lopsidedness and recent star
 formation is poorly understood and so at the present time this effect
 cannot be accounted for.

\section{Conclusions and Future Work}
We have determined the median $\langle \tilde{A}_1\rangle$ in our
 sample of early type disk galaxies to be $0.11$.  We have also
 determined that $20\%$ of the galaxies in our sample have $\langle
 \tilde{A}_1\rangle \geq 0.19$.  By looking in the R-band and by
 selecting galaxies with typically low star formation rates, we have
 picked a compromise between efficiency and the minimization of the
 contributions from young (i.e. non phase-mixed), bright stars and
 hence we are primarily viewing actual azimuthal variations in the
 stellar mass distributions of the early type disks in our sample.

By comparing the value of $\langle \tilde{A}_1\rangle$ which is lower
 than that for $50\%$ and $20\%$ of our galaxy sample, to that
 measured for late type spirals by ZR97 (0.13 and 0.19 respectively),
 we can address the issue of whether the asymmetries seen in ZR97 are
 also caused by variations in the stellar mass distribution of late
 type disk galaxies.  If asymmetric star formation were important in
 creating asymmetric light distributions, we would expect that late
 type disk galaxies would have a higher incidence of lopsidedness than
 early type spirals.  Since this is not the case, we are led to
 conclude that lopsidedness in disk galaxies of all types reflects
 primarily variations in the stellar mass distributions.  Our data
 then show that one-fifth of {\it all} disk galaxies have $\langle
 \tilde{A}_1\rangle\geq0.19$.

The volume correction discussed in section 4.2 affects our measurement
 of the frequency of lopsidedness and the possibility of systematic
 brightening of lopsided galaxies via an increased star formation rate
 needs to be explored in more detail.  Comparison of the recent star
 formation history of lopsided galaxies to that of symmetric galaxies
 is needed to study this effect, and is also needed to help better
 establish a cause of lopsidedness.  A study of this nature is being
 carried out this spring by the authors.

Further N-Body simulations also need to be carried out to determine if
 the types and magnitudes of asymmetries seen in our samples can be
 reproduced by weak interactions as well as by minor mergers.  It is
 also necessary to better determine if isolated phenomena
 (e.g. instabilities) result in asymmetries similar in morphology and
 magnitude to those we have measured.

\acknowledgments
Greg Rudnick and H.-W. Rix would like to thank Robert Kennicutt for
valuable discussions.  We would also like to thank James Pizagno for
assisting with our observing program.

\clearpage
\begin{deluxetable}{llllllll}
\tablewidth{0pt}
\tablecaption{Image~Sample}
\tablehead{\colhead{${\rm~object~name}$} & \colhead{${\rm~T}$} & \colhead{${\rm~D}_{25}\tablenotemark{a}$} & \colhead{${\rm~v}_{rec}$} & \colhead{${\rm~b/a}\tablenotemark{b}$} & \colhead{${\rm~R}_{exp}$} & \colhead{${\rm~R}_{max}/{\rm~R}_{exp}$} & \colhead{$\langle {\rm~\tilde{A}}_1\rangle $}\\
\colhead{$\phm{0}$} & \colhead{$\phm{0} $} & \colhead{$[ \arcmin ]$} & \colhead{$[km/s]$} & \colhead{$\phm{0} $} & \colhead{$[ \arcsec ]$} & \colhead{$\phm{0} $} & \colhead{$\phm{0} $}}
\startdata
$IC~520$ & $2.0$ & $1.95$ & $3528$ & $0.79$ & $13.0$ & $2.95$ & $0.088\pm.126$ \nl
$NGC~0023$ & $1.0$ & $2.09$ & $4635$ & $0.64$ & $14.0$ & $2.93$ & $0.049\pm0.039$ \nl
$NGC~0828$ & $1.0$ & $2.88$ & $5181$ & $0.77$ & $15.5$ & $4.68$ & $0.12\pm0.045$ \nl
$NGC~2342$ & $0.0$ & $1.38$ & $5209$ & $0.93$ & $14.6$ & $2.62$ & $0.56\pm0.066$ \nl
$NGC~2460$ & $1.0$ & $2.45$ & $1442$ & $0.76$ & $7.4$ & $4.72$ & $0.045\pm0.038$ \nl
$NGC~2551$ & $0.2$ & $1.66$ & $2263$ & $0.67$ & $12.0$ & $4.25$ & $0.063\pm0.069$ \nl
$NGC~2554\tablenotemark{c}$ & $0.0$ & $1.50$ & $4126$ & $0.74$ & $14.0$ & $3.43$ & $0.092\pm0.032$ \nl
$NGC~2599$ & $1.0$ & $1.86$ & $4690$ & $0.89$ & $16.5$ & $2.87$ & $0.124\pm0.031$ \nl
$NGC~2681$ & $0.0$ & $3.63$ & $692$ & $0.91$ & $32.5$ & $2.92$ & $0.040\pm0.037$ \nl
$NGC~2775\tablenotemark{c}$ & $2.0$ & $1.63$ & $1340$ & $0.77$ & $17.4$ & $5.08$ & $0.093\pm0.026$ \nl
$NGC~2782$ & $1.0$ & $3.47$ & $2562$ & $0.74$ & $20.4$ & $2.87$ & $0.19\pm0.098$ \nl
$NGC~2855$ & $0.0$ & $2.45$ & $1910$ & $0.89$ & $9.0$ & $4.70$ & $0.11\pm0.11$ \nl
$NGC~2993\tablenotemark{c}$ & $1.0$ & $1.13$ & $2227$ & $0.69$ & $9.4$ & $4.40$ & $0.162\pm0.055$ \nl
$NGC~3277$ & $2.0$ & $1.95$ & $1460$ & $0.89$ & $8.2$ & $4.37$ & $0.046\pm0.048$ \nl
$NGC~3415$ & $-0.5$ & $2.09$ & $3177$ & $0.63$ & $9.8$ & $4.01$ & $0.038\pm0.037$ \nl
$NGC~3442$ & $1.0$ & $0.62$ & $1724$ & $0.76$ & $3.5$ & $4.11$ & $0.286\pm0.047$ \nl
$NGC~3504$ & $2.0$ & $2.69$ & $1518$ & $0.77$ & $19.6$ & $3.86$ & $0.106\pm0.046$ \nl
$NGC~3611$ & $1.0$ & $2.09$ & $1754$ & $0.81$ & $12.2$ & $3.77$ & $0.118\pm0.043$ \nl
$NGC~3682$ & $0.0$ & $1.66$ & $1543$ & $0.66$ & $7.2$ & $4.65$ & $0.145\pm0.060$ \nl
$NGC~3720$ & $0.9$ & $0.95$ & $5958$ & $0.91$ & $5.4$ & $4.38$ & $0.122\pm0.060$ \nl
$NGC~3732\tablenotemark{c}$ & $0.0$ & $1.09$ & $1682$ & $0.95$ & $6.53$ & $3.62$ & $0.169\pm0.026$ \nl
$NGC~3884$ & $0.0$ & $2.09$ & $6869$ & $0.64$ & $13.7$ & $3.37$ & $0.213\pm0.210$ \nl
$NGC~4245$ & $0.0$ & $2.88$ & $890$ & $0.76$ & $18.2$ & $3.42$ & $0.026\pm0.038$ \nl
$NGC~4277\tablenotemark{c}$ & $0.0$ & $1.02$ & $2516$ & $0.83$ & $9.0$ & $3.26$ & $0.014\pm0.028$ \nl
$NGC~4314$ & $1.0$ & $4.17$ & $963$ & $0.89$ & $11.5$ & $5.40$ & $0.010\pm0.035$ \nl
$NGC~4369\tablenotemark{c}$ & $1.0$ & $1.32$ & $988$ & $0.98$ & $8.6$ & $5.41$ & $0.238\pm0.068$ \nl
$NGC~4378$ & $1.0$ & $2.88$ & $2563$ & $0.93$ & $12.2$ & $3.46$ & $0.027\pm0.037$ \nl
$NGC~4384\tablenotemark{c}$ & $1.0$ & $1.11$ & $2400$ & $0.77$ & $6.7$ & $4.48$ & $0.153\pm0.040$ \nl
$NGC~4415$ & $0.0$ & $1.35$ & $825$ & $0.89$ & $7.0$ & $4.54$ & $0.040\pm0.992$ \nl
$NGC~4421$ & $0.0$ & $2.69$ & $1692$ & $0.76$ & $17.0$ & $3.01$ & $0.014\pm0.037$ \nl
$NGC~4454\tablenotemark{c}$ & $0.0$ & $1.30$ & $2373$ & $0.85$ & $14.6$ & $3.39$ & $0.080\pm0.038$ \nl
$NGC~4457$ & $0.0$ & $2.69$ & $738$ & $0.85$ & $22.8$ & $2.72$ & $0.113\pm0.042$ \nl
$NGC~4464$ & $0.0$ & $1.07$ & $1255$ & $0.76$ & $5.0$ & $4.27$ & $0.006\pm0.036$ \nl
$NGC~4470$ & $1.0$ & $1.29$ & $2358$ & $0.72$ & $10.6$ & $3.39$ & $0.494\pm0.075$ \nl
$NGC~4492$ & $1.0$ & $1.70$ & $1772$ & $0.93$ & $10.4$ & $4.01$ & $0.115\pm0.143$ \nl
$NGC~4566$ & $0.0$ & $1.26$ & $5290$ & $0.71$ & $12.4$ & $2.54$ & $0.151\pm0.081$ \nl
$NGC~4643$ & $0.0$ & $3.09$ & $1399$ & $0.74$ & $4.9$ & $10.40$ & $0.006\pm0.035$ \nl
$NGC~4665$ & $0.0$ & $3.80$ & $785$ & $0.83$ & $8.2$ & $7.57$ & $0.008\pm0.035$ \nl
$NGC~4670$ & $0.0$ & $1.44$ & $1112$ & $0.76$ & $7.5$ & $5.61$ & $0.348\pm0.052$ \nl
$NGC~4691$ & $0.0$ & $2.82$ & $1108$ & $0.81$ & $13.1$ & $4.74$ & $0.448\pm0.057$ \nl
$NGC~4763$ & $1.0$ & $1.51$ & $4160$ & $0.71$ & $8.4$ & $4.14$ & $0.091\pm0.050$ \nl
$NGC~4765\tablenotemark{c}$ & $0.0$ & $1.06$ & $706$ & $0.72$ & $5.0$ & $5.09$ & $0.120\pm0.026$ \nl
$NGC~4778$ & $-1.0$ & $0.71$ & $4355$ & $0.67$ & $4.9$ & $2.55$ & $0.008\pm0.037$ \nl
$NGC~4795$ & $0.5$ & $1.86$ & $2684$ & $0.85$ & $18.8$ & $2.73$ & $0.254\pm0.104$ \nl
$NGC~5548\tablenotemark{c}$ & $0.0$ & $1.16$ & $5026$ & $0.89$ & $9.0$ & $3.25$ & $0.192\pm0.050$ \nl
$NGC~5614$ & $2.0$ & $2.45$ & $3872$ & $0.83$ & $10.1$ & $3.93$ & $0.117\pm0.096$ \nl
$NGC~5691$ & $1.0$ & $1.86$ & $1846$ & $0.76$ & $9.8$ & $5.23$ & $0.357\pm0.139$ \nl
$NGC~5695$ & $0.0$ & $1.55$ & $4206$ & $0.71$ & $9.4$ & $2.51$ & $0.109\pm0.044$ \nl
$NGC~5701\tablenotemark{c}$ & $0.0$ & $1.63$ & $1556$ & $0.95$ & $11.0$ & $4.22$ & $0.012\pm0.025$ \nl
$NGC~5915$ & $2.0$ & $1.74$ & $2245$ & $0.72$ & $5.0$ & $3.48$ & $0.204\pm0.109$ \nl
$NGC~5963$ & $0.0$ & $3.31$ & $770$ & $0.77$ & $7.8$ & $5.40$ & $0.076\pm0.065$ \nl
$NGC~6012$ & $2.0$ & $2.09$ & $1988$ & $0.72$ & $11.4$ & $3.70$ & $0.032\pm0.042$ \nl
$NGC~6646$ & $1.0$ & $1.23$ & $5764$ & $0.85$ & $7.4$ & $4.43$ & $0.012\pm0.038$ \nl
$NGC~6962$ & $2.0$ & $2.88$ & $4254$ & $0.79$ & $10.4$ & $3.69$ & $0.053\pm0.037$ \nl
\label{tbl:table.9.3.in}
\enddata
\tablenotetext{a}{the diameter of the galaxy out to the isophote of surface brightness 25 mag/arcsec$^2$}
\tablenotetext{b}{the ratio of the minor to major axis as defined in the RC3 catalogue
}\tablenotetext{c}{The last three columns are the averages for multiple images
}\end{deluxetable}

\clearpage
\begin{deluxetable}{lllll}
\tablewidth{0pt}
\tablecaption{Characteristics~of~Multicolor~Sample}
\tablehead{\colhead{${\rm~object~name}$} & \colhead{${\rm~T}$} & \colhead{${\rm~D}_{25}\tablenotemark{a}$} & \colhead{${\rm~v}_{rec}$} & \colhead{${\rm~b/a}\tablenotemark{b}$}\\
\colhead{$\phm{ } $} & \colhead{$\phm{ } $} & \colhead{$[ ' ]$} & \colhead{$[km/s]$} & \colhead{$\phm{ }$}}
\startdata
$IC~1269$ & $4.0$ & $1.70$ & $6116$ & $.74$ \nl
$NGC~5600$ & $5.0$ & $1.44$ & $2319$ & $.95$ \nl
$NGC~6555$ & $5.0$ & $2.00$ & $2225$ & $.78$ \nl
\label{tbl:table2.9.4.in}
\enddata
\tablenotetext{a}{the diameter of the galaxy out to the isophote of surface brightness 25 mag/arcsec$^2$}
\tablenotetext{b}{the ratio of the minor to major axis as defined in the RC3 catalogue
}\end{deluxetable}

{}

\clearpage
\begin{figure}
\epsscale{1.00}
\plotone{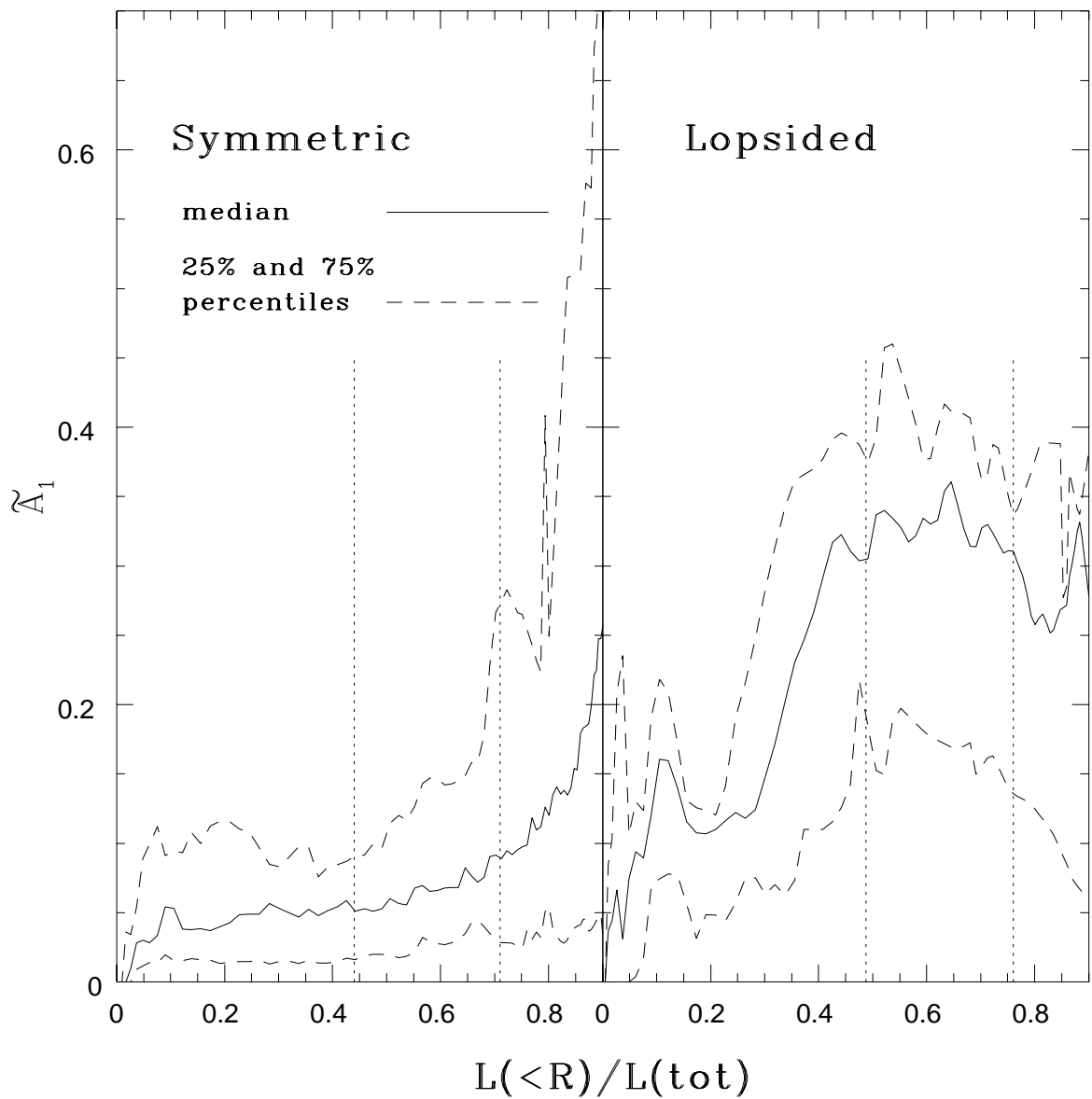}
\caption{
The median $\tilde{A}_1(L(\leq R)/L(tot))$ profiles for the $20\%$ most
 lopsided galaxies in our sample and the $80\%$ least lopsided for
 whom $L(\leq R)/L(tot)\geq0.71$ (R$_{max}\geq$2.5R$_{exp}$.)  The dotted
 lines contain the central 50$\%$ of the points.  The vertical dotted
 lines indicate the radial range over which $\langle
 \tilde{A}_1\rangle$ was determined.  }
\end{figure}
\clearpage

\begin{figure}
\epsscale{1.00}
\plotone{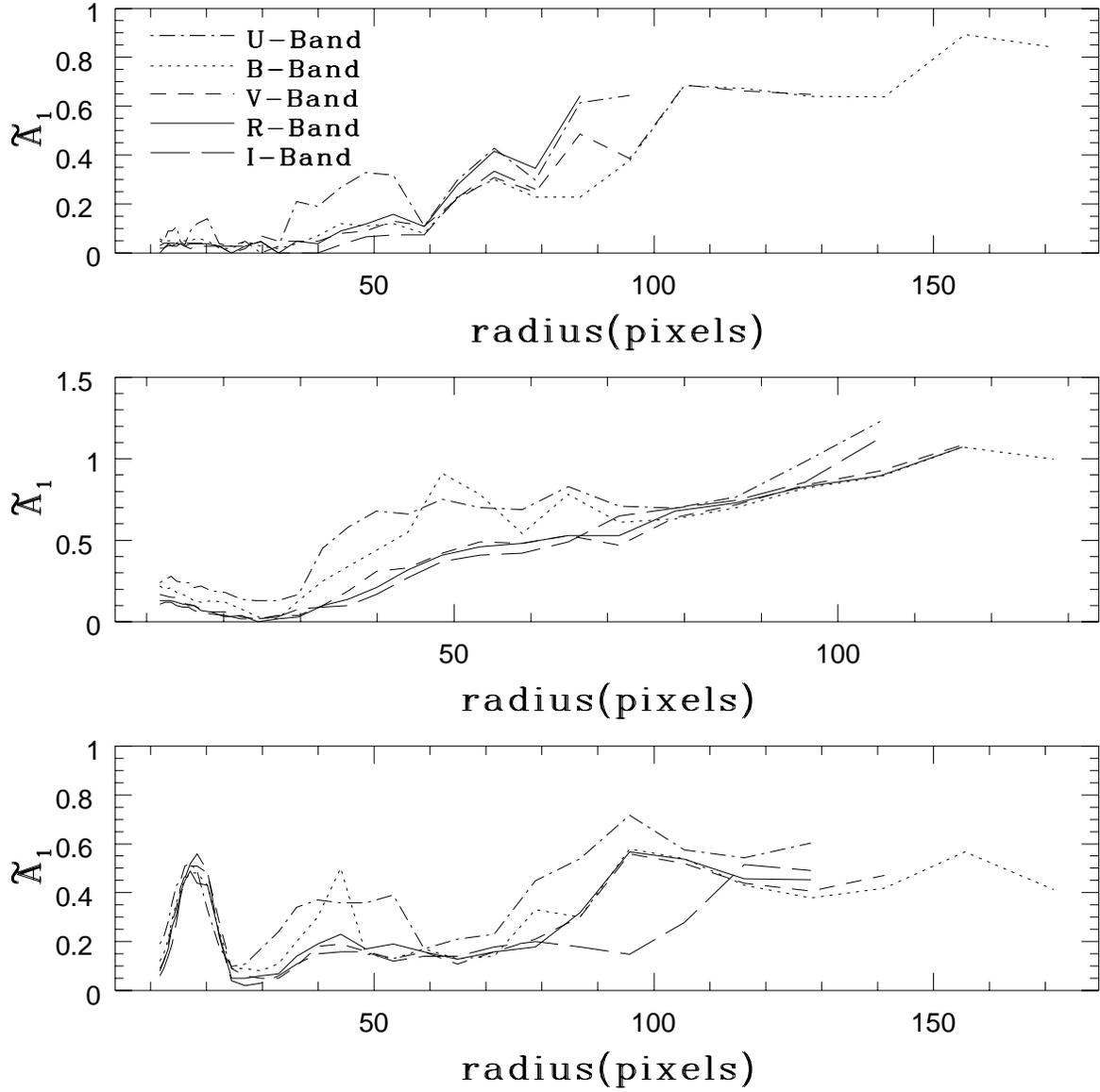}
\caption{
The UBVRI $\tilde{A}_1(R)$ profiles for three late type spiral galaxies: IC~
1269~(top), NGC~5600~(middle) and NGC~6555~(top).  U and B-band images
exhibit deviations from the mean profile, but V,R and I-band images
are generally congruent.}
\end{figure}
\clearpage

\begin{figure}
\epsscale{1.00}
\plotone{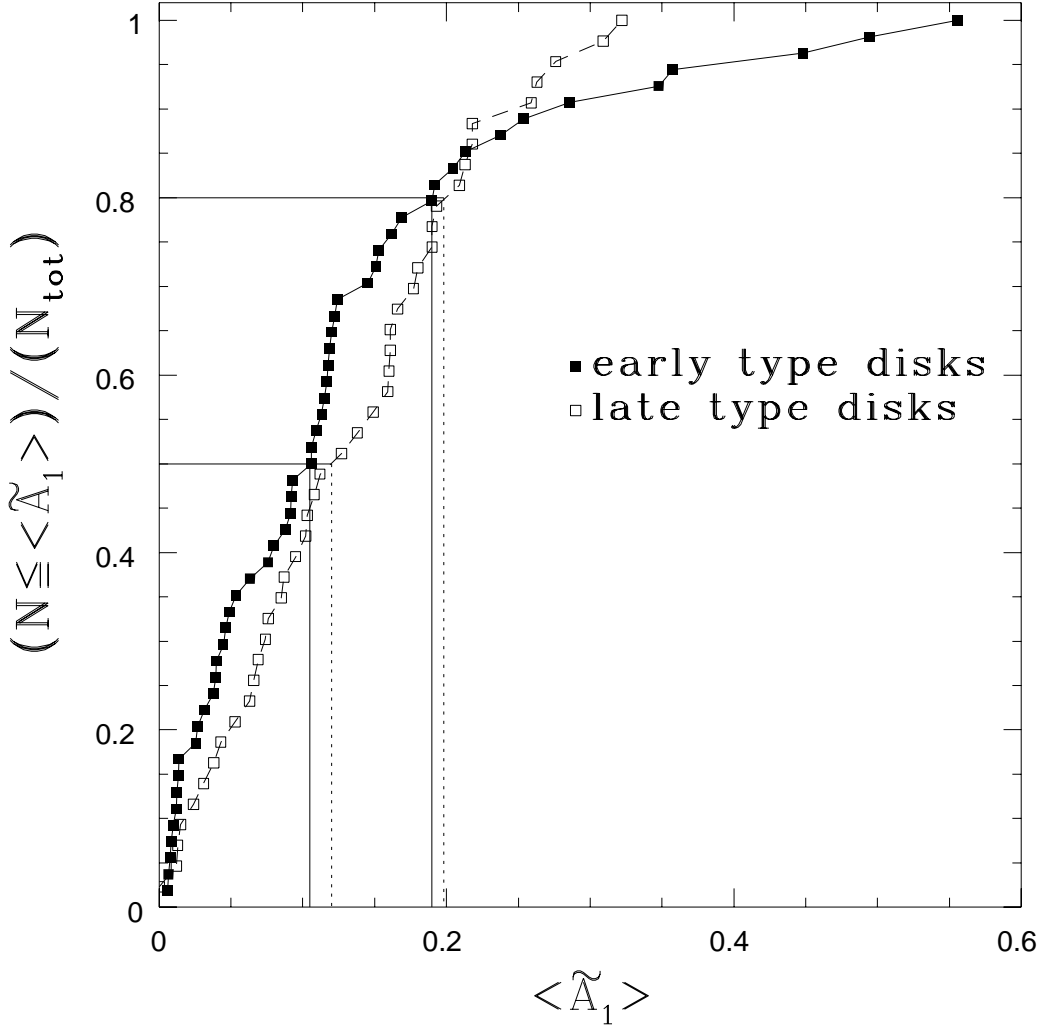}
\caption{
The cumulative histograms of $\langle \tilde{A}_1\rangle $ for the
 current sample and for the sample of ZR97.  The very small
 differences between $\langle A_1\rangle $ and $\langle
 \tilde{A}_1\rangle $ in ZR97 do not affect our comparison of the two
 samples.  The y-axis is the fraction of galaxies in the sample with
 $\langle \tilde{A}_1\rangle \leq$ a certain amount.  There are 43
 Late Type disk and 54 early type disk galaxies.  Note that the
 current sample has 3 galaxies with significantly higher $\langle
 \tilde{A}_1\rangle $ than ZR97.  We account for the ZR97 star
 subtraction methods by subtracting the offset (0.03) from all of
 their data points. }
\end{figure}

\end{document}